\documentstyle{elsart}

\begin{document}

\begin{frontmatter}
\title{Short Distance Behavior of (2+1)-dimensional QCD}
\author{Holger Gies}
\address{Institut f\"ur theoretische Physik der Universit\"at 
       T\"ubingen\\D-72076 T\"ubingen\\ Germany\\
       E-mail address: holger.gies@uni-tuebingen.de}

\begin{abstract}
  Within the framework of semiclassical QCD approximations the short
  distance behavior of two static color charges in (2+1)-dimensional
  QCD is discussed. 
  A classical linearization of the field equations is exhibited and
  leads to analytical results producing the static potential. Beyond
  the dominant classical part proportional to $\ln{\lambda R}$, QCD
  contributions of order $R^{1/2}$ and $R$ are found.
\end{abstract}

\begin{keyword}
 PACS 11.15.K
\end{keyword}
\end{frontmatter}

\section{ Introduction}
 During the past years many attempts have been made to approximate the
theory of quantum chromodynamics (QCD) by abelianized, classical
models. Especially Adler and Piran \cite{1a,1b} managed to analyze the 
potential of two static, opposite color charges of (3+1)-dimensional
QCD (QCD$_4$) with regard to short and long distance behavior using
such methods. In particular, they found a confining potential plus
correction terms.\\
The idea consists in describing the vacuum and its properties as a
dielectric medium, arising as a consequence of the quantum
fluctuations of Yang-Mills fields. One handles the quantum
fluctuations with the help of effective Lagrangians whose application
leads to nonlinear Maxwell equations of electrostatics.\\
In most cases, these equations are not exactly soluble, but at least
it is possible to obtain the asymptotic behavior of the theory, e.g.,
corrections to classical field theory.\\
In this work we study (2+1)-dimensional quantum chromodynamics
(QCD$_3$) which has been shown to be a confining theory by Frenkel
and Silva Fo. \cite{2}. Unlike the latter, we focus on the short
distance behavior of two static QCD$_3$ color charges, separated by a 
distance of $R$, and their static potential.\\
In section 2 we introduce the effective action approach to QCD$_3$.
Section 3 answers the question of how the dielectric model is
correctly approximated by linear electrostatics employing a formalism
developed by Lehmann and Wu \cite{3}. Finally, in section 4 we show
how calculation of QCD contributions to the static potential is
nevertheless possible, and we obtain the classical part proportional
to $\ln \lambda R$ followed by terms of order $R^{1/2}$ and $R$.\\

\section{Effective action in QCD$_3$}
In 4-dimensional space the static potential is defined by:

\begin{equation}
  V_{\mathrm{static}}=-{\mathrm{extremum}}_{\varphi}\left[ \int {\d} 
  ^{3}x({\cal L}(\nabla \varphi )-\varphi j_{0})\right] +\triangle 
  V_{\mathrm{c}} \label{1}
\end{equation}

with Lagrangian density $\cal L$, potential function $\varphi $,
current density $j_{0}$ and\\
Coulomb counter terms $\triangle
V_{\mathrm{c}}$ to remove self-energies.\\ 
We calculated the 1-loop
effective Lagrangian of pure QCD$_{3}$ by summing over normal modes of 
a constant color field by means of, for example, the gauge invariant
$\zeta $-function regularization method. Our findings agree with
Trottier's result \cite{9}:

\begin{equation}
  {\cal L}_{\mathrm{eff}}(\nabla \varphi )=\frac{1}{2}(\nabla \varphi
  )^{2} \left[ 1-\frac{4}{3} \left( \frac{(\nabla \varphi
  )^{2}}{\kappa ^{2}} \right) ^{-1/4} \right] \label{5}
\end{equation}

where $\kappa ^{2}>0$ denotes the minimum of ${\cal
L}_{\mathrm{eff}}$. For example, in pure QCD$_{3}$ without fermions:  

\begin{equation}
  \kappa ^{1/2}=g^{3/2}\frac{3}{4\pi } \left[ 1+\frac{1}{\sqrt{2} } 
  -\frac{\zeta (3/2)}{\sqrt{2}\ 4\pi } \right] \label{6}
\end{equation}

where $\zeta (x)$: Riemannian Zeta-function. Besides, $\kappa $ is not
a renormalization group invariant, because there is simply no
renormalization group due to the superrenormalizability of the
model.\\ 
By inserting (\ref{5}) into (\ref{1}) , we establish quantum
fluctuations as a mean field background and define the {\em leading 
root\/} modell of QCD$_{3}$ analogous with the {\em leading log\/}
modell of QCD$_4$ \cite{4a,4b}.\\

\section{Short distance approximation}
It is obvious that the effective Lagrangian (\ref{5}) is already
decomposed into the classical part $\propto (\nabla \varphi)^2$ and a
part containing the QCD corrections $\propto
[(\nabla\varphi)^2]^{\frac{3}{4}}$. Hence, ${\cal L}_{\mathrm eff}$
naturally reduces to the classical Lagrangian because of increasing
field strength while $R$ decreases.  In order to treat this statement
more quantitatively we have to write down the field equations of our 
dielectric model. Employing the Euler-Lagrange equations and ${\cal
L}_{\mathrm{eff}}$ (\ref{5}), yields:

\begin{eqnarray}
 \mbox{\boldmath$\nabla \cdot D$} =j_{0} & & \epsilon^{ij} E^{j}=0
 \label{8a} \\
 \mbox{\boldmath$D$} =\epsilon
 \mbox{\boldmath$E$} & \epsilon =1- \sqrt{\frac{\kappa }{E} }&
 E=|\mbox{\boldmath$E$} |\label{8b} \\
 & j_{0}=Q(\delta
 ^{2}(|\mbox{\boldmath$r_{1}$}|)-\delta
 ^{2}(|\mbox{\boldmath$r_{2}$}|)) & \label{8c} \\
 \mbox{\boldmath$r_{1}$} =(x-a,y) & \mbox{\boldmath$r_{2}$} =(x+a,y) &
 R=2a \label{8d}
\end{eqnarray}

The difficulty of imposing boundary conditions on (\ref{8a}) is
conveniently handled via introducing a manifestly flux conserving form
\cite{6}. In this method, $\mbox{\boldmath$D$}$ is expressed as a
function of the electric flux through a curve $C$ which intersects the
charge axes at a point $x_{\mathrm{s}}>a$ as shown the following
figure (fig.1):

\begin{figure}[h] 
\begin{center}
\unitlength0.9cm
\begin{picture}(10,5)(-5,-2.5)
 \put(-5,0){\vector(1,0){10}} \put(4.8,-0.3){$x$}
 \put(0,-2.5){\vector(0,1){5}} \put(0.2,2.2){$y$} 
 \put(3,0){\circle*{0.2}} \put(3,0.3){$Q$} \put(3,-0.5){$a$}
 \put(-3,0){\circle*{0.2}} \put(-3,0.3){$\bar{Q} $}
                             \put(-3.3,-0.5){$-a$}
 \put(2,0){\oval(4,4)[r]}
 \put(4.1,-0.3){$x_{\mathrm{s}}$}
 \put(2,2){\circle*{0.1}} \put(4,-2.2){$C$} \put(0.9,2){$(x,y)$}
 \put(2,-2){\circle*{0.1}}
\end{picture}
\end{center}
fig.1
\end{figure}

This has to be done carefully in two dimensions due to the inequality
of line and surface integrals although lines and surfaces are similar
objects. Explicit calculation with the aid of an arbitrary
parametrization of $C$ shows that the flux definition $\Phi
=\int\limits_{C} \mbox{\boldmath$D \cdot dn$} $ is always satisfied by
the choice:

\begin{equation}
   \mbox{\boldmath$D$} =\frac{1}{2} \left(
         \begin{array}{c}
                \partial _{y} \Phi \\
               - \partial _{x} \Phi
         \end{array}
         \right) \label{9}
\end{equation}         

Now, it is possible to incorporate boundary conditions by imposing
 them on the flux function $\Phi (x,y)$:

\begin{equation}
 \Phi (y=0)= \left\{
   \begin{array}{cc}
      Q \mbox{  if} & |x|<a \\
      0 \mbox{  if} & |x|>a
   \end{array}
   \right. \label{10}
\end{equation}

\begin{center} 
 $ \Phi \rightarrow 0 \mbox{  if  } x^{2}+y^{2} \rightarrow \infty $
\end{center}

The dynamical equation for $\Phi$ now comes from the second equation
of (\ref{8a}):
$\epsilon^{ij} E^{j}=0$, also expressing $E$ in terms of $\Phi $ by
inverting (\ref{8b}):

\begin{equation}
 E=f(D) =\frac{\kappa }{2} +\sqrt{\frac{\kappa ^{2}}{4} +\kappa D} +D
 \label{12}
\end{equation}

where $D=|\mbox{\boldmath$D$}|=\half \sqrt{(\partial _{x}\Phi )^{2}
+(\partial _{y}\Phi )^{2}} $.\\
Calculating $\epsilon^{ij} E^{j}=0$ with the help of (\ref{12}) we
obtain the exact quasilinear, second order differential equation of
elliptic type (as Lehmann and Wu found in QCD$_{4}$):

\begin{equation}
 0=[\Phi _{x}^{2} +({\scriptstyle \frac{g(D)}{D} } +1)\Phi
  _{y}^{2}]\Phi _{xx}+ [({\scriptstyle \frac{g(D)}{D} } +1)\Phi
  _{x}^{2}+\Phi _{y}^{2}]\Phi _{yy}- 2{\scriptstyle \frac{g(D)}{D}  }
  \Phi _{x}\Phi _{y} \Phi _{xy} \label{13} 
\end{equation} 

where $\Phi _{x}$ is a short form of $\partial _{x}\Phi $ and $g(D)=
\frac{f(D)}{f'(D)} -D $ denotes a coefficient function.\\
Expanding $g(D)$ in terms of $D$ near the sources when $D\gg \kappa $,
yields: 

\begin{equation}
 g(D)=\frac{1}{2} \sqrt{\kappa D} +\frac{\kappa }{4} +{\cal O}
 (D^{-1/2}) \label{14}
\end{equation}

Thus $\frac{g(D)}{D} $ is of order ${\cal O}(D^{-1/2})$; therefore we
are allowed to omit it for short distances obtaining, not
surprisingly, Laplace's equation as approximation of (\ref{13}):

\begin{equation}
 0=(\partial ^{2}_{x} +\partial ^{2}_{y})\Phi \label{15}
\end{equation}

which has the well-known solution satisfying the boundary conditions
(\ref{10}) :

\begin{equation}
  \Phi =\Phi _{\mathrm{cl}}=\frac{Q}{\pi} \left[ \arctan{\left(
   \frac{y}{x-a} \right) } -\arctan{\left( \frac{y}{x+a} \right) }
   \right] \label{16}
\end{equation}

Furthermore we find:

\begin{equation}
 \mbox{\boldmath$D=D$}_{\mathrm{cl}}=\frac{Q}{2\pi }\left(
  \frac{\mbox{\boldmath$\hat{r} _{1}$}}{r_{1}}
  -\frac{\mbox{\boldmath$\hat{r} _{2}$}}{r_{2}} \right) 
  \label{17}
\end{equation}

and the potential function:

\begin{equation}
 \varphi =\varphi _{\mathrm{cl}} =-\frac{Q}{2\pi }
  \ln{\frac{r_{1}}{r_{2}} } =-\frac{Q}{2\pi }
  \ln{\frac{\sqrt{(x-a)^{2}+y^{2}} }{\sqrt{(x+a)^{2}+y^{2} } } }
  \label{18}
\end{equation}

To summarize, we come to the hardly surprising but important
conclusion that the classical linearization is appropriate for
approximating the short distance behavior of two static color charges.

\section{The static potential}   
In order to finally compute the static potential including QCD$_{3}$
contributions we employ another formula of $V_{\mathrm{static}}$ that
is shown to be equal to (\ref{1}) \cite{1a,1b}:

\begin{equation}
 V_{\mathrm{static}}=\int {\d } ^{2}x \int\limits_{0}^{D} {\d }
 \tilde{D}\  f(\tilde{D} ) \label{19}
\end{equation}

where $f(D)=E(D)$. Inserting (\ref{12}) and expanding $(D\gg \kappa)$
leads to: 

\begin{eqnarray}
 V_{\mathrm{static}} & = & \int {\d } ^{2}x \left[ \frac{\kappa }{2}
 D+ \frac{2}{3\kappa } \sqrt{\frac{\kappa ^{2}}{4} +D\kappa } ^{\,3} 
 -\frac{\kappa ^{2}}{12} +\frac{1}{2} D^{2}\right] \nonumber\\
 & = &\int {\d} ^{2}x \left[ \frac{1}{2} D^{2}+\frac{2}{3}
 \sqrt{\kappa} D^{3/2}+\frac{\kappa }{2} D+{\cal O}(D^{1/2})\right]
 \label{20} \\ 
 & = & V_{\mathrm{static}}^{D^{2}}+V_{\mathrm{static}}^{D^{3/2}}+
 V_{\mathrm{static}}^{D}+\dots \nonumber
\end{eqnarray}

Of course, $V_{\mathrm{static}}^{D^{2}}$ is simply the classical
potential and is well known from electrostatics or might be calculated
directly by partial integration and use of Poisson's equation with due
consideration of Coulomb counter terms:

\begin{equation}
 V_{\mathrm{static}}^{D^{2}}=\frac{Q^{2}}{2\pi } \ln{\lambda R}
 \label{21} 
\end{equation}

where $\lambda $ denotes a mass parameter without physical
significance.\\
In the following it proves to be useful to integrate
$V_{\mathrm{static}}^{D}$ first. Note that $D=\frac{1}{2}\sqrt{(
\partial_{x}\Phi )^{2}+  (\partial _{y}\Phi )^{2}} $ can be written
as: 

\begin{equation}
 D=\frac{Qa}{\pi } \left[ \left( (x-a)^{2}+y^{2}\right) \left(
 (x+a)^{2}+y^{2} \right) \right] ^{-1/2} \label{22} 
\end{equation}

We are able to perform the $y$-integration without restrictions using
the symmetry of our system \cite{7}:\\

\begin{eqnarray}
 V_{\mathrm{static}}^{D} & = & \frac{\kappa }{2} \int {\d } ^{2}x\
 D=2\kappa \int\limits_{0}^{\infty} {\d } x\int\limits_{0}^{\infty}
 {\d } y\ D\nonumber\\ & = & \frac{1}{\pi} \kappa QR
 \int\limits_{0}^{\infty } {\d } k\ K(k) \quad ,\quad k=\frac{x}{a}
 \label{23}
\end{eqnarray}

where $K(k)$ denotes the complete elliptic integral of the first kind.
Now, the $k$-integration does not make sense, because $K(k)$ is well
defined for $k^{2}<1$ (or $|x|<a$) only. But there is no reason for
concern; as we obtain from the analysis of long distance behavior
\cite{2,8} , the free boundary of confinement intersects the $x$-axes
at $x_{\mathrm{b}}=\pm a$ (in QCD$_{3}$ as well as in QCD$_{4}$).
Simply because $D$ vanishes outside the boundary the remaining
integration interval is part of the defining interval of $K(k)$. We
find :

\begin{equation}
 V_{\mathrm{static}}^{D}\rightarrow \frac{1}{\pi} \kappa QR
 \int\limits_{0}^{1} {\d } k\ K(k)= \frac{2G}{\pi} \kappa QR=0.58\dots
 \kappa QR \label{24}
\end{equation}

where $G$ denotes Catalan's constant. It should not come as a surprise
that we obtain a term $\propto R$ as in the case of long distances due
to the appearance of a term $\propto D$, since it is responsible for
the creation of linear confinement both in QCD$_{3}$ and QCD$_{4}$. Of
course, here it is of less importance.  Besides, the limit of
integration removes divergent self-energy terms in an elegant way.\\
The leading order contribution to the classical potential is found by
the integration of $V_{\mathrm{static}}^{D^{3/2}}$ which has to be
performed with great care. First of all we insert the potential
function $\varphi $, integrate by parts and use Poisson's equation:

\begin{eqnarray}
 V_{\mathrm{static}}^{D^{3/2}} & = &\frac{2}{3} \sqrt{\kappa } \int
 {\d } ^{2}x\ D^{3/2}= \frac{2}{3} \sqrt{\kappa } \int {\d } ^{2}x
 \frac{\nabla \varphi \cdot \nabla \varphi }{\sqrt{D} } \nonumber\\ 
 & = & \frac{2}{3} \sqrt{\kappa} \left\{ \int {\d} ^{2}x\ \nabla\cdot
 \left( \varphi \frac{\nabla \varphi}{\sqrt{D} } \right) +\frac{1}{2}
 \int {\d } ^{2}x \frac{\varphi j_{0}}{\sqrt{D} } \right\} \label{25}
\end{eqnarray}

The second integral completely disappears because the integrands
behave as\\ $\lim_{p\to 0} p\ln{p} $. For the remaining integral we
use the same integration volume as in the case of
$V_{\mathrm{static}}^{D}$ which takes account of the confining
boundary and automatically removes divergent self-energies. Next we
take care of the specific properties of Gauss' law in two dimensions
and choose the following path of integration with the normal unit
vectors pointing outwards (fig.2):

\begin{figure}[h]
\begin{center}
\unitlength0.35cm
\begin{picture}(16,14)(-8,-7)
 \put(-5.7,-0.8){$-a$}
 \put(4.3,-0.8){$a$}
 \put(0.2,-6.8){$-\infty $}
 \put(-1.4,6.3){$\infty $}
 \put(0,-7){\vector(0,1){14}}
 \put(-8,0){\vector(1,0){16}}
 \put(-4,-6){\vector(1,0){5}}
 \put(4,-5){\vector(0,1){7}}
 \put(-4,6){\vector(1,0){4}}
 \put(-4,-5){\vector(0,1){7}}
 \put(1,-6){\line(1,0){3}}
 \put(4,2){\line(0,1){3}}
 \put(0,6){\line(1,0){4}}
 \put(-4,2){\line(0,1){3}}
 \multiput(-4,-6)(0,0.2){6}{\circle*{0.05}}
 \multiput(4,-6)(0,0.2){6}{\circle*{0.05}}
 \multiput(-4,5)(0,0.2){6}{\circle*{0.05}}
 \multiput(4,5)(0,0.2){6}{\circle*{0.05}}
 \put(-1,-6){\vector(0,-1){1}}
 \put(1,6){\vector(0,1){1}}
 \put(4,1){\vector(1,0){1}}
 \put(-4,1){\vector(-1,0){1}}
 \put(7.2,-0.7){$x$}
\end{picture}
\end{center}
fig.2
\end{figure}

We obtain:

\begin{eqnarray}
  V_{\mathrm{static}}^{D^{3/2}} & =\frac{2}{3} \sqrt{\kappa } &
 \left\{ \int\limits_{-a}^{a} {\d } x\lim_{y\to -\infty }
 (-\mbox{\boldmath$\hat{y}$}) 
 \cdot \mbox{\boldmath$W$} + \int\limits_{-a}^{a} {\d } x\lim_{y\to
 \infty } (\mbox{\boldmath$\hat{y}$}) \cdot \mbox{\boldmath$W$}
 \right. \nonumber\\ 
 &&\left. \mbox{}+\int\limits_{-\infty}^{\infty} {\d} y\lim_{x\to -a}  
 (-\mbox{\boldmath$\hat{x}$}) \cdot \mbox{\boldmath$W$}
 +\int\limits_{-\infty }^{\infty 
 } {\d } y\lim_{x\to a} (\mbox{\boldmath$\hat{x}$}) \cdot
 \mbox{\boldmath$W$} \right\} \label{26}
\end{eqnarray}

\begin{displaymath}
 \mbox{where\/}\ \mbox{\boldmath$W$} :=\frac{\varphi \nabla \varphi
 }{\sqrt{D} } 
 =\frac{1}{4} \sqrt{\frac{Q^{3}}{a\pi ^{3}} } \ln \left(
 {\frac{r_{1}}{r_{2}}} \right) \left[
 \frac{\mbox{\boldmath$r_{1}$ } \sqrt{r_{2}} }{\sqrt{r_{1}^{3}} }
 -\frac{\mbox{\boldmath$r_{2}$ } \sqrt{r_{1}} }{\sqrt{r_{2}^{3}} }
 \right]
\end{displaymath}

The $x$-integrals vanish due to $\mbox{\boldmath$W$}$ approaching zero
at infinity.  There are no divergencies nor are there any analytical
problems in the remaining $y$-integrations; hence we find
$V_{\mathrm{static}}^{D^{3/2}}$ evaluating to:

\begin{equation}
 V_{\mathrm{static}}^{D^{3/2}}=\frac{\sqrt{2} }{3} \left( \frac{\pi
 ^{2}}{2} - \frac{\Psi'(1/4)}{4} \right) \sqrt{\frac{\kappa Q^{3}}{\pi
 ^{3}} } \hspace{0.5cm} R^{1/2} \label{27}
\end{equation}

where $\Psi'(x)$ denotes the derivative of the psi function
and $\Psi'(1/4) \simeq $ 17.1973.\\ 
Therefore the potential of two static color charges $(Q-\bar{Q} )$ of 
QCD$_{3}$ in the case of small separations is found to be:

\begin{eqnarray}
 V_{\mathrm{static}} & = & \frac{Q^{2}}{2\pi } \ln{\lambda R}
 +\frac{\sqrt{2} }{3} \left( \frac{\pi ^{2}}{2} -\frac{\Psi' \left(
 \frac{1}{4}\right)}{4} \right) \sqrt{\frac{\kappa Q^{3}}{\pi ^{3}} }
 \hspace{0.2cm} R^{1/2} +\left(\frac{2G}{\pi }+\dots \right)
 \kappa QR \nonumber\\
 & = & \frac{Q^{2}}{2\pi } \ln{\lambda R} +0.054.. 
 \sqrt{\kappa Q^{3}} R^{1/2} +(0.583..+\dots) \kappa QR \label{28}
\end{eqnarray}

In addition to the dominant classical potential, there are subdominant
contributions behaving like $R^{1/2}$, $R$, \dots\ vanishing as $R$
approaches zero. In the last correction term, it is indicated that the
numerical factor will be modified due to quantum corrections of the
flux function itself. The magnitude of this contribution lies outside
the various approximations applied to reach (\ref{28}). Detailed
analysis shows that the first correction term $\propto R^{1/2}$ is not
changed!  Equation (\ref{28}) should be read side by side with Adler's
formula (40) of reference \cite{4b}.

\section{Conclusion}

Despite important differences in physical and analytical details, the
global structure of QCD$_{3}$ seems to be comparable to that of
QCD$_{4}$ with reference to quasiclassical approximations, as was
pointed out in more detail by Cornwall \cite{5}. Although QCD$_{3}$
does not provide renormalization group methods, a Callan-Symanzik
$\beta $-function or a mass-scale dependent running coupling, it is
possible within the framework of Adler and Piran to study the
asymptotic freedom of a confining theory.\\ It is interesting that
quantum contributions result from the employment of solutions of
purely classical differential equations.  Here it proved appropriate
not to approximate a solution to an exact differential equation but to
approximate the differential equation itself.\\ In this way the short
distance approximation of the static potential was determined
including nonclassical contributions of order $R^{1/2}$ and $R$.  Even
with reservations due to the relevance of low-dimensional field
theories the calculations may lend some insight into the mechanisms of
classical approximations of real QCD.\\
\bigskip\\ 

\begin{ack}
I would like to thank Prof. W. Dittrich for suggesting the problem and
for carefully reading the manuscript.\\
\end{ack}

\end{document}